\documentclass{elsart}  
\usepackage[dvips]{graphicx}
\usepackage{amssymb}

\begin{document}

\begin{frontmatter}

\title{Determination of the basic timescale in kinetic Monte Carlo
  simulations by comparison with cyclic-voltammetry experiments}
\author[add1,add2]{I. Abou Hamad}
\author[add1,add2]{P.A. Rikvold\corauthref{cor1}}
\ead{rikvold@csit.fsu.edu}
\author[add2,add3]{G. Brown}
\corauth[cor1]{Corresponding author. 
Address: $^a$ Tel.: +1-850-644-6011; Fax: +1-850-644-0098}
\address[add1]{Center for Materials Research and Technology and Department
of Physics, Florida State University, Tallahassee, FL 32306-4350, USA}
\address[add2]{School of Computational Science, Florida State University,
\\Tallahassee, FL 32306-4120, USA}
\address[add3]{Center for Computational Sciences, Oak Ridge National 
Laboratory, \\
Oak Ridge, TN 37831-6164, USA}

\begin{abstract}
While kinetic Monte Carlo simulations can provide long-time
simulations of the dynamics of physical and chemical systems, it is
not yet possible in general to identify the 
inverse Monte Carlo attempt frequency with a physical timescale 
in any but the simplest systems. Here we demonstrate such an
identification by comparing simulations with experimental data. Using
a dynamic lattice-gas model for the electrosorption of Br on Ag(100), we
measure the scan-rate dependence of the separation between positive- and
negative-going peaks in cyclic voltammetry and compare
simulated and experimental peak separations. By adjusting the Monte Carlo
attempt frequency, good agreement between simulated and experimental
peak separations is achieved. 
It is also found that the uniqueness of the 
determination depends on the relative values 
of the adsorption/desorption and diffusion free-energy barriers.
\end{abstract}

\end{frontmatter}

\section{Introduction}
\label{sec:I}

At present, kinetic Monte Carlo (KMC) simulation is virtually the only
computational method that enables numerical study of the dynamics of
physical and chemical systems on macroscopically relevant timescales, 
anywhere from microseconds to millions of 
years \cite{Kolesik,Brown,Combe,Auer,Novotny}.
However, the method is essentially a stochastic approximation to 
underlying
classical or quantum mechanical processes on finer
space and time scales~\cite{NowakPRL:00,Faraday}. It thus suffers from
the problem that the basic MC timescale is often difficult to
relate to an underlying physical timescale. While
such a timescale could, in principle, be calculated from comparison
with calculations at these finer                
scales~\cite{NowakPRL:00}, this is in practice rarely possible for
electrochemical systems.
Due to the complexity of the interactions with the solution, ab-initio
methods can construct a reasonable horizontal corrugation
potential~\cite{Faraday} but give limited knowledge about the shape
of the potential in the direction perpendicular to the surface.
Moreover, water has effects both in terms of damping and in the shape
of the adsorption/desorption free-energy barrier, 
which probably corresponds to a
reconstruction of the solvation shell. While analytic macroscopic theories
can be derived in terms of MC parameters \cite{BerthierJEAC:04-1and2}, such
theories cannot directly predict the values of those parameters. 
A rough estimate of the overall reaction rate constant can be obtained 
by applying standard electrochemical techniques \cite{Srinivasan:66}
to the experimental data. Yet, the overall reaction rate gives little
information, if any, about the free-energy barrier heights for the
different processes of adsorption/desorption and surface diffusion.

Here we present an alternative approach: comparison of KMC
results with time-dependent experimental results. In particular, we
compare KMC simulations of a lattice-gas model with experimental
results for cyclic-voltammetry (CV) studies of the electrosorption of
Br on single-crystal Ag(100) surfaces. The lattice-gas model represents the
long-lived configurations of adsorbed Br, and an 
MC step corresponds to an attempt at 
hopping across a saddle point in the free-energy landscape 
to a new configuration~\cite{Faraday}.

The Br/Ag(100) system has a phase transition 
(in the two-dimensional Ising universality class) 
between a disordered phase at more negative potentials and
an ordered $\rm{c} (2 \times 2)$ phase at more positive
potentials~\cite{OCKO:BR/AG}. The phase transition is associated with
a divergence of the coverage fluctuations, corresponding to a peak in
the cyclic voltammogram. The same phase transition has also been observed
for Cl/Ag($100$) in electrochemical~\cite{Hamad1} and ultra-high
cavuum (UHV) environments~\cite{Taylor,Hwang}.

Recent static and kinetic MC studies
have been used to investigate the phase ordering and
disordering mechanisms in cyclic-voltammetry (CV) and sudden
potential-step experiments for halide adsorption on Ag(100) \cite{A,B}. 
As the CV scan rate is increased, the system is driven further from
equilibrium. As a result, there is a widening of the separation between the
positions of the CV peak for the positive-going and the negative-going
scan of the electrode potential (in experiments) or the
electrochemical potential (in simulations). Experimental peak
separations have been measured for different sweep rates of the
electrode potential~\cite{Jia}. When these experimental
values, with time measured in seconds, are compared to the 
simulations, with time measured in Monte Carlo steps per site (MCSS),
a physical time can be associated with the inverse MC
attempt frequency. This was previously attempted by 
Mitchell \textit{et al.}~\cite{MitchellLandau},
but simulations at that time could not achieve 
peak separations within the
experimental range. Due to the increase in computer power and to a new
mean-field enhanced simulation method that we developed
for calculating long-range interactions~\cite{Hamad1}, it is now possible
to simulate peak separations well within the experimental range.

\section{Lattice-gas Model}
\label{sec:LGM}

We employ a LG model similar to that used by Koper
\cite{KOPER:HALIDE,KOPE98C} and Mitchell,
\textit{et al.} \cite{A,B}. The Br ions adsorb on four-fold hollow
sites of the Ag(100) surface~\cite{Faraday,Sanwu1}. The model is 
defined by the grand-canonical effective Hamiltonian,
\begin{equation}
{\mathcal{H}} = - \sum_{i<j} \phi_{ij} c_{i} c_{j} - \overline{\mu} 
\sum_{i=1}^{N}c_{i} \; ,
\label{eq:H}
\end{equation}
where $\sum_{i<j}$ is a sum over all pairs of sites, $ \phi_{ij} $ are
the lateral interactions between particles on the $i$th and
$j$th lattice sites measured in~meV/pair, $ \overline{ \mu } $ is the
electrochemical potential measured in meV/particle, and $N=L^2$ is the
total number of lattice sites. The local
occupation variables $ c_{i} $ can take the values 1 or 0, depending
on whether site $ i $ is occupied by an ion (1) or empty (0).

The simulations were performed on $L\times L$ square lattices, using
periodic boundary conditions to reduce finite-size effects.  The
interaction constants $ \phi_{ij} $ between ions on sites $i$ and $j$
a distance $r_{ij}$ apart (measured in units of the Ag(100) lattice
spacing, $a=2.889$~{\AA} \cite{OCKO:BR/AG}) are given by
\begin{equation}
\phi_{ij} = - \infty \times  \delta_{r_{ij}, 1} 
+ \frac{2^{3/2} \phi_{\rm nnn} }{r_{ij}^{3}} \; , 
\end{equation}
where $\delta_{r_{ij}, 1}$ is a Kronecker delta with 
the infinitely negative value for $r_{ij}=1$ indicating nearest-neighbor
exclusion. Negative values of $\phi_{\rm nnn}$ denote long-range repulsion. The
interactions for large $r_{ij}$ are most likely of electrostatic dipole-dipole
nature~\cite{Hamad3}, but may also have a component mediated by
the Ag substrate~\cite{EinsteinLANG:91} that may not be uniformly repulsive.

Assuming a sufficient concentration of counterions in the electrolyte,
the electrochemical potential $\overline{ \mu }$ is, in the dilute-solution
approximation, related to the bulk ionic concentration $C$ and
the electrode potential $E$ (measured in mV) as
\begin{equation} 
\overline{\mu} = \overline{\mu}_{0} + k_{\mathrm B} T \ln \frac{C}{C_{0}} 
- e \gamma E \; ,
\label{eq:mbar}
\end{equation}
where $ \overline{\mu}_{0} $ is an arbitrary constant, $C_{0}$ is a
reference concentration (here $1~\rm mM$), $k_{\rm B} T$ is
Boltzmann's constant times the temperature, and $ e $ is the
elementary charge unit. The electrosorption
valency $\gamma$~\cite{C,Schmickler} (here assumed
constant~\cite{Hamad1,Hamad3})
corresponds to the fraction of the ionic charge transferred during
the adsorption process. We use the sign convention
that $ \overline{ \mu } > 0$ favors adsorption.

The adsorbate coverage is defined as
$\theta = N^{-1} \displaystyle \sum_{i=1}^{N}{c_{i}}$.
It can
be experimentally obtained by standard electrochemical methods, as well
as from the integer-order peaks in surface X-ray scattering (SXS)
data~\cite{OCKO:BR/AG,Th.Wandlowski}. The derivative of the
coverage with respect to the electrochemical potential, 
$\rm{d}\theta/\rm{d}\overline{\mu}$, is proportional to the current
density in an experimental CV as described in section~\ref{sec:Results}.

We have previously estimated the 
next-nearest-neighbor lateral interaction energy ($\phi_{\rm nnn}\approx-21
$~meV) and the electrosorption valency ($\gamma\approx-0.72$), based on fits
of simulated equilibrium adsorption isotherms to experimental 
data~\cite{Hamad1}. Remarkably, despite the very different environments this
value of $\phi_{\rm nnn}$, estimated from the shape of the adsorption isotherms,
is consistent to within $10\%$ with the one obtained from the temperature
dependence of the critical coverage for Cl/Ag($100$) in UHV~\cite{Taylor,Hwang}.
In the next section we detail the kinetic MC approach.

\section{Kinetic Monte Carlo Method}
\label{sec:Methods}
Kinetic Monte Carlo simulations were performed on systems with
$L=128$, and the absence of significant finite-size effects was
verified by additional simulations for $L=64$ and 256.
The kinetic MC simulation proceeds as follows. We randomly select a
lattice site $i$, and depending on the occupation $c_{i}$, different
moves are attempted. If $c_{i}=0$,
only adsorption is attempted, while if $c_{i}=1$, $9$
different moves are proposed: desorption, diffusion to each of the $4$
nearest-neighbor sites, and diffusion to each of the $4$
next-nearest-neighbor sites. Next, a weighted
list for accepting each of these moves is constructed using
Eq.~(\ref{eq:P}) below, to calculate the probabilities
$R\rm{(F|I)}$ of the individual moves between the initial state $\rm I$
and final state $\rm F$. The probability for
the system to stay in the initial configuration is consequently 
$R\rm{(I|I)}=1-$$\Sigma_{\rm F\neq I}R\rm{(F|I)}$~\cite{B}.

Using a thermally activated, stochastic
barrier-hopping picture, the energy of the transition state for a
microscopic change from an initial state $\rm I$ to a final state
$\rm F$ is approximated by the symmetric Butler-Volmer
formula~\cite{Brown,Kang,Buendia}
\begin{equation}
U_{\rm T_{\lambda}}=\frac{U_{\rm I}+U_{\rm F}}{2}+\Delta_{\lambda} \; ,
\end{equation}
where $U_{\rm I}$ and $U_{\rm F}$ are the energies of the initial and final
states, respectively, $\rm T_{\lambda}$ is the transition state for process
$\lambda$, and $\Delta_{\lambda}$ is a ``bare'' barrier associated with process
$\lambda$, which here can be one of nearest-neighbor diffusion
($\Delta_{\rm nn}$), next-nearest-neighbor diffusion ($\Delta_{\rm nnn}$), or
adsorption/desorption ($\Delta_{\rm a/d}$). 

The probability for a particle to make a transition from state $\rm I$ to
state $\rm F$ is approximated by the one-step Arrhenius
rate~\cite{Brown,Kang,Buendia}
\begin{equation}
\mathcal{R}(\rm F|\rm I)=  \nu \exp \left(-\frac{\Delta_{\lambda}}{k_{\rm B}T}\right)
\exp\left(-\frac{U_{\rm F}-U_{\rm I}}{2k_{\rm B}T}\right),\label{eq:P}
\end{equation}
where $\nu$ is the attempt frequency, which sets the overall
timescale for the simulation. 
This formalism assumes that all processes $\lambda$
have the same MC attempt frequency of $1$ MCSS$^{-1}$. Although this
assumption may not be true in general~\cite{BowlerJCP:94}, because 
different processes involve different potential wells and thus different
vibrational frequencies, it can be used to find an effective 
timescale conversion between MC time and physical time, given
the MC attempt frequencies. The MC time unit, one  
MCSS, corresponds to the attempted update of $N$ randomly chosen sites.
For a simulation with scan rate $\rho$ (in meV/MCSS), $\overline{\mu}$
is incremented by ($\rho \times 1$~MCSS) meV every MCSS 
until it reaches its
final value, and then decremented back to its initial value. 

Since the value of $U_{\rm F}-U_{\rm I}$ determines $\mathcal{R}$ in
Eq.~(\ref{eq:P}), the approximations made to calculate the
large-$r$ contributions to the pair sum in Eq.~(\ref{eq:H}) are
important. In our simulations, to calculate the energy changes 
we included the exact contributions for particle separations up to $r_{ij}=3$,
while using a mean-field approximation for larger separations~\cite{Hamad1}. 
\footnote{Omission of the mean-field part, which could be viewed as
a simple-minded way to introduce electrostatic screening, would increase
the value of $\phi_{\rm nnn}$ that best fits the equilibrium isotherms
by $5-10\%$~\cite{Hamad1,Hamad3}.}
\section{Results}
\label{sec:Results}

\begin{figure}
\centerline{\includegraphics[angle=0,width=6in]{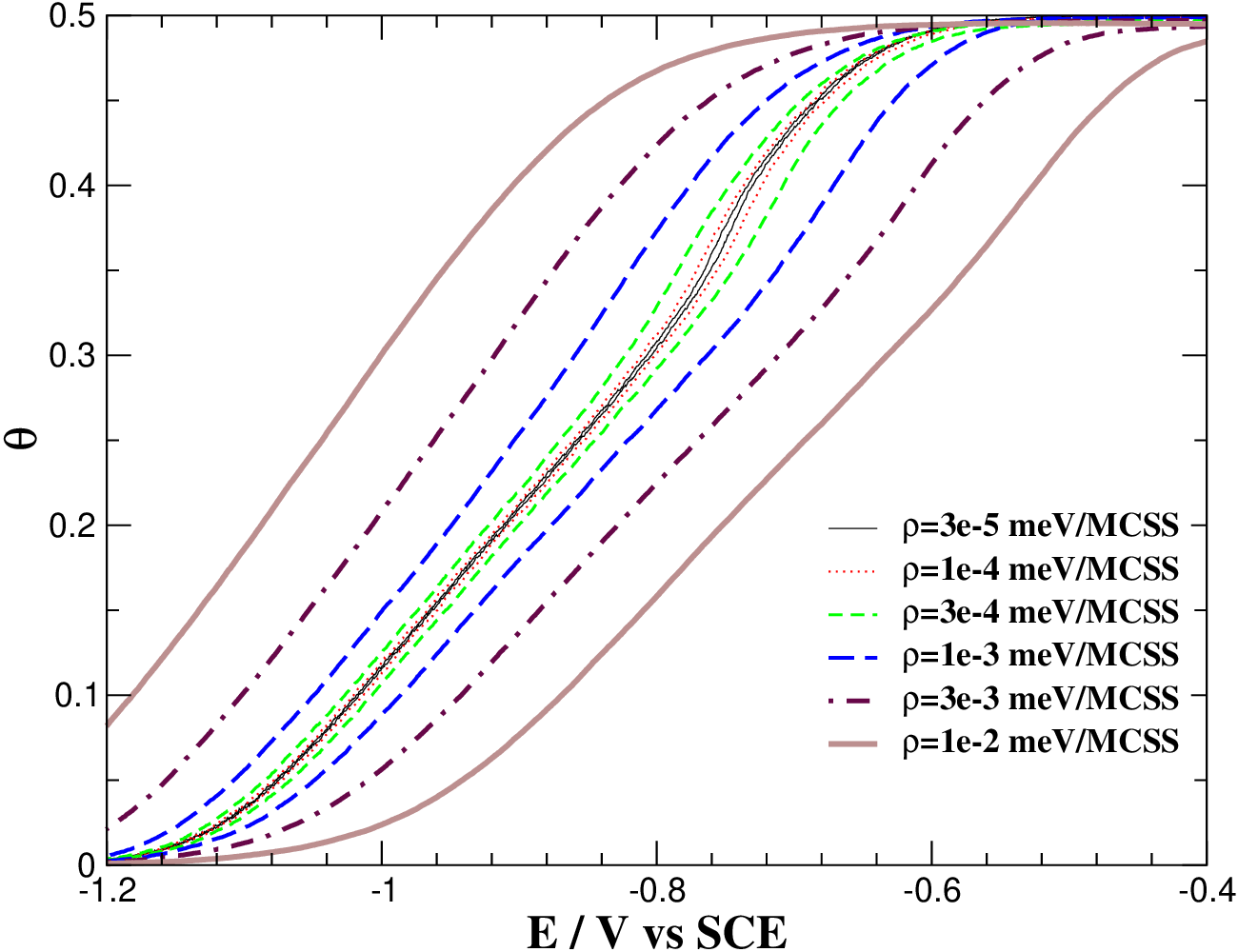}}
\caption[]{The coverage $\theta$ as a function of the electrode
  potential $E$ for
  $\Delta_{\rm a/d}=300 $~meV and different scan rates.
\label{fig:covs}}
\end{figure}

Starting at an initial potential of $\overline{\mu}=-200 $~meV and
room temperature ($\beta=1/k_{\rm B}T=0.04~\rm{meV^{-1}}$), the MC
process discussed in Sec.~\ref{sec:Methods} was repeated until
$\overline{\mu}$ reached $+600$~meV, and then decremented at the same
rate back to $-200$~meV.
The coverage isotherms were computed using scan rates
ranging over four decades, from $\rho = 3\times 10^{-5}$ to 0.1~meV/MCSS, 
and for $\Delta_{\rm a/d}=150$, 175, 200, 250, 300, 350, and
$400$~meV. The other barriers, $\Delta_{\rm nnn}=200$~meV and
$\Delta_{\rm nn}=100$~meV, were kept constant at the values determined
by comparison with density-functional theory calculations~\cite{Faraday,Sanwu1}.
The hysteresis loops for 
$\theta$ as a function of $E$ are shown for $\Delta_{\rm
  a/d}=300$~meV in Fig.~\ref{fig:covs} for different scan rates. Each
loop was averaged over eight independent simulation runs.

\begin{figure}
\centerline{\includegraphics[angle=0,width=6in]{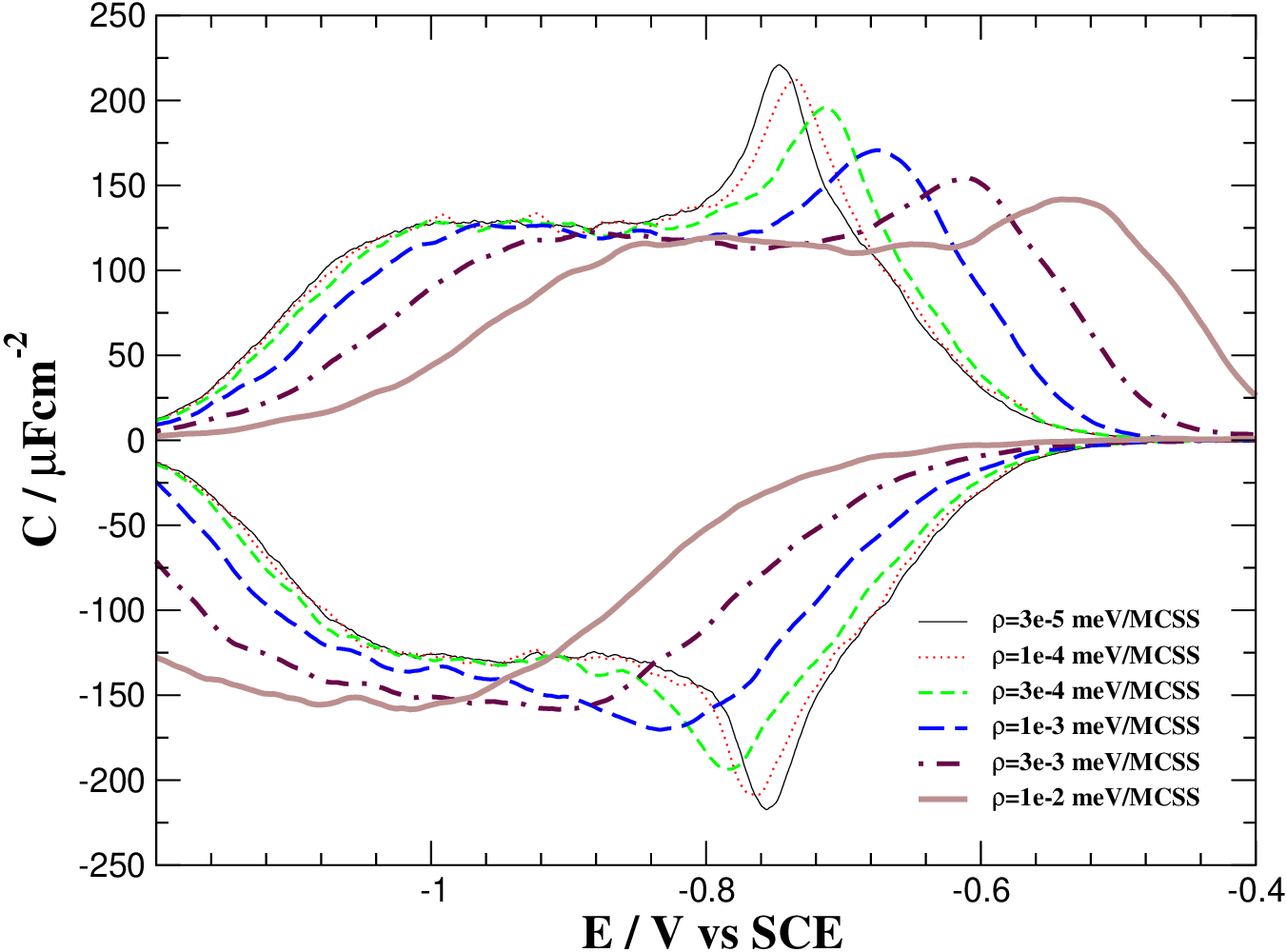}}
\caption[]{Simulated adsorption capacitance cyclic voltammograms
for $\Delta_{\rm a/d}=300 $~meV and different scan rates.
\label{fig:CVs}}
\end{figure}

Next, the Savitzky-Golay method~\cite{SavGol,Recipes} with a second-order
polynomial and a window of 51 points was used to obtain the
smoothed numerical derivative $\rm{d}\theta/\rm{d}\overline{\mu}$, which is
proportional to the simulated CV current density~$j$~\cite{B}:
\begin{equation}
  j=\frac{\gamma^2 e^2}{A_{\rm s}}\frac{\rm{d}\theta}{\rm{d}\overline{\mu}}\frac{{\rm d}E}{{\rm d} t}, 
\end{equation}
where $A_{s}$ is the area of one
adsorption site. Consequently, the differential adsorption capacitance
per unit area becomes
\begin{equation}
C=\frac{j}{{\rm d}E / {\rm d} t}=\frac{\gamma^2 e^2}{A_{\rm s}}\frac{\rm{d}\theta}{\rm{d}\overline{\mu}}\; .
\end{equation}
As seen in~Fig.~\ref{fig:CVs}, the difference between the positions of
the positive-going and negative-going peaks depends on the scan rate. This
difference is due to two physical effects. First, jamming 
due to the nearest-neighbor exclusion, which is only slowly alleviated
by lateral diffusion \cite{B}, 
causes the coverage to lag behind its equilibrium value. Second, 
critical slowing down also causes the coverage to equilibrate slowly
near the phase transition at the CV peak \cite{B}.\footnote{
If $j$ is plotted vs $\theta$, rather than vs $E$ or $\overline{\mu}$, 
all the peaks fall in the range $0.37 \le \theta \le 0.40$ for 
positive-going scans and $0.33 \le \theta \le 0.37$ for negative-going
scans. This is consistent with the
values of the hard-square jamming coverage and critical coverage \cite{Taylor,Hwang,B}.
}

\begin{figure}
\centerline{\includegraphics[angle=0,width=6in]{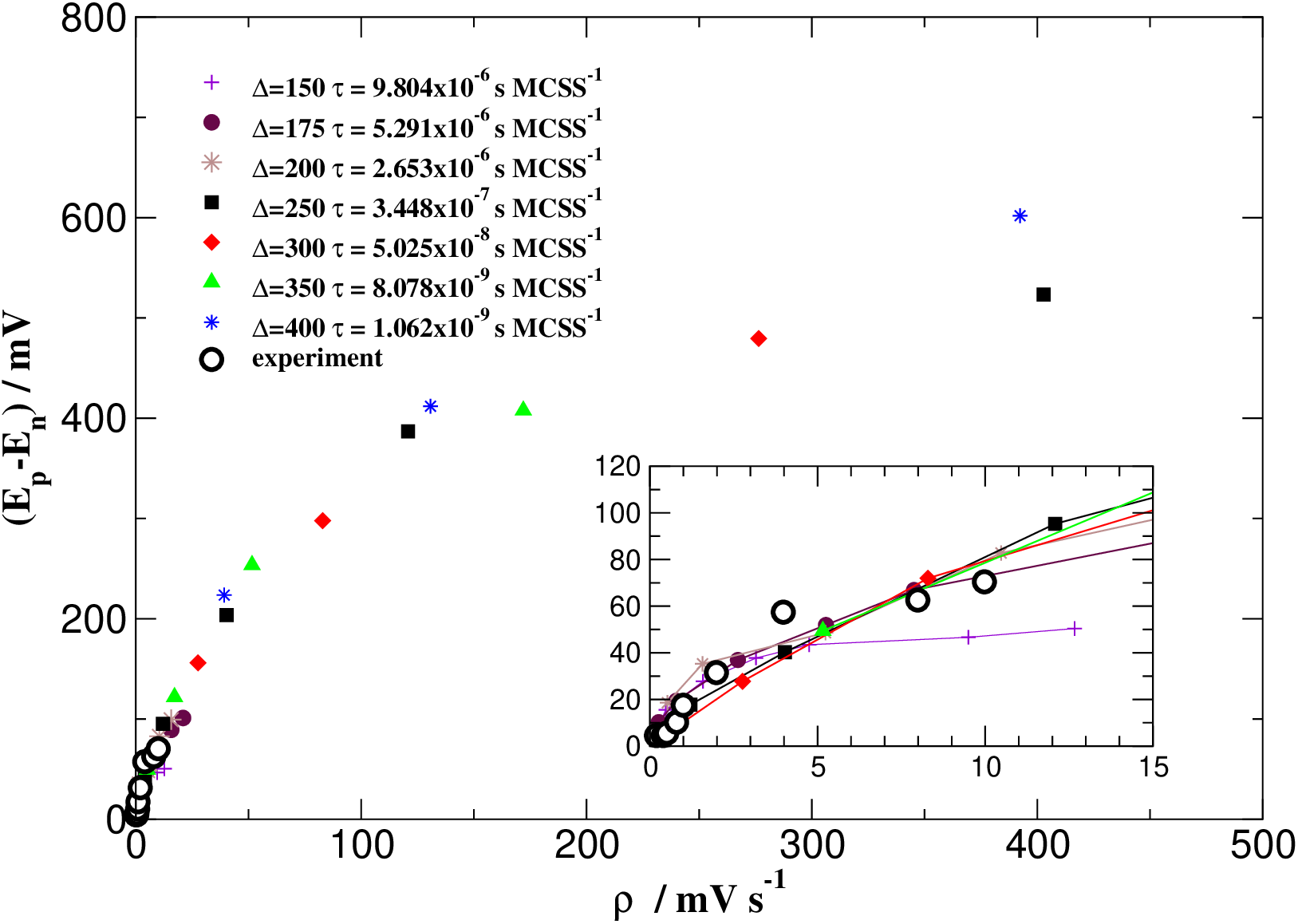}}
\caption[]{Fits of kinetic MC simulations to experimental peak
  separations using different values for the adsorption-desorption
  barrier, $\Delta_{\rm a/d}$. Zooming into the experimental data
  range (inset) suggests that the best-fit value for $\Delta_{\rm a/d}$
  is $175$~meV, corresponding to $\tau\approx5.3\times10^{-6}$ s. 
  The lines are guides to the eye.
\label{fig:peaksep}}
\end{figure}

By fitting the separation of the positive-going scan peak ($E_p$)
and the negative-going scan peak ($E_n$) of the simulated CVs to experimental
data for $E_{p}-E_{n}$ vs d$E$/d$t$~\cite{Jia}, we can extract
a physical time $\tau$ corresponding to 1 MCSS (the inverse MC
attempt frequency, $\tau =1/\nu$) for each value of the adsorption/desorption
free-energy barrier height $\Delta_{\rm{a}/\rm{d}}$.
A $\chi^2$ which measures the square of the horizontal
differences between the experimental curves and the simulated peak separations, 
multiplied by trial values of $\tau$, was calculated. The
value of $\tau$ which minimized $\chi^2$ was taken to be the best-fit value.
See~Fig.~\ref{fig:peaksep}. For
$\Delta_{\rm{a/d}} \leq 350$~meV,
the simulations were fit to the experimental data points, while for
$\Delta_{\rm{a/d}}=400$~meV the simulations were fit to the best-fit simulated
data points for $\Delta_{\rm{a/d}}=300$~meV since the peak separations
attainable for $\Delta_{\rm{a/d}}=400$~meV are not within the experimental 
data range. The fits for most values of $\Delta_{\rm{a/d}}$ coincide to within the
accuracy of the experimental data and our statistics. Yet, focusing on
the experimental data range (inset in Fig.~\ref{fig:peaksep}),
suggests that $\Delta_{\rm{a/d}}=175\;\rm{meV}$ and
$\tau=5.3\times10^{-6}$ s fits best to the experimental data. This
distinction relies mainly on the two experimental data points corresponding
to the highest scan rates and
thus would be better founded if there were more experimental data
in that range. In addition, since the simulated curves for
$\Delta_{\rm a/d} > 200$~meV practically coincide, it is only for
$\Delta_{\rm{a/d}}\leq 200$~meV that a clear distinction
can be made among the different $\Delta_{\rm{a/d}}$ values, even with
more accurate experimental data.

\begin{figure}
\centerline{\includegraphics[angle=0,width=6in]{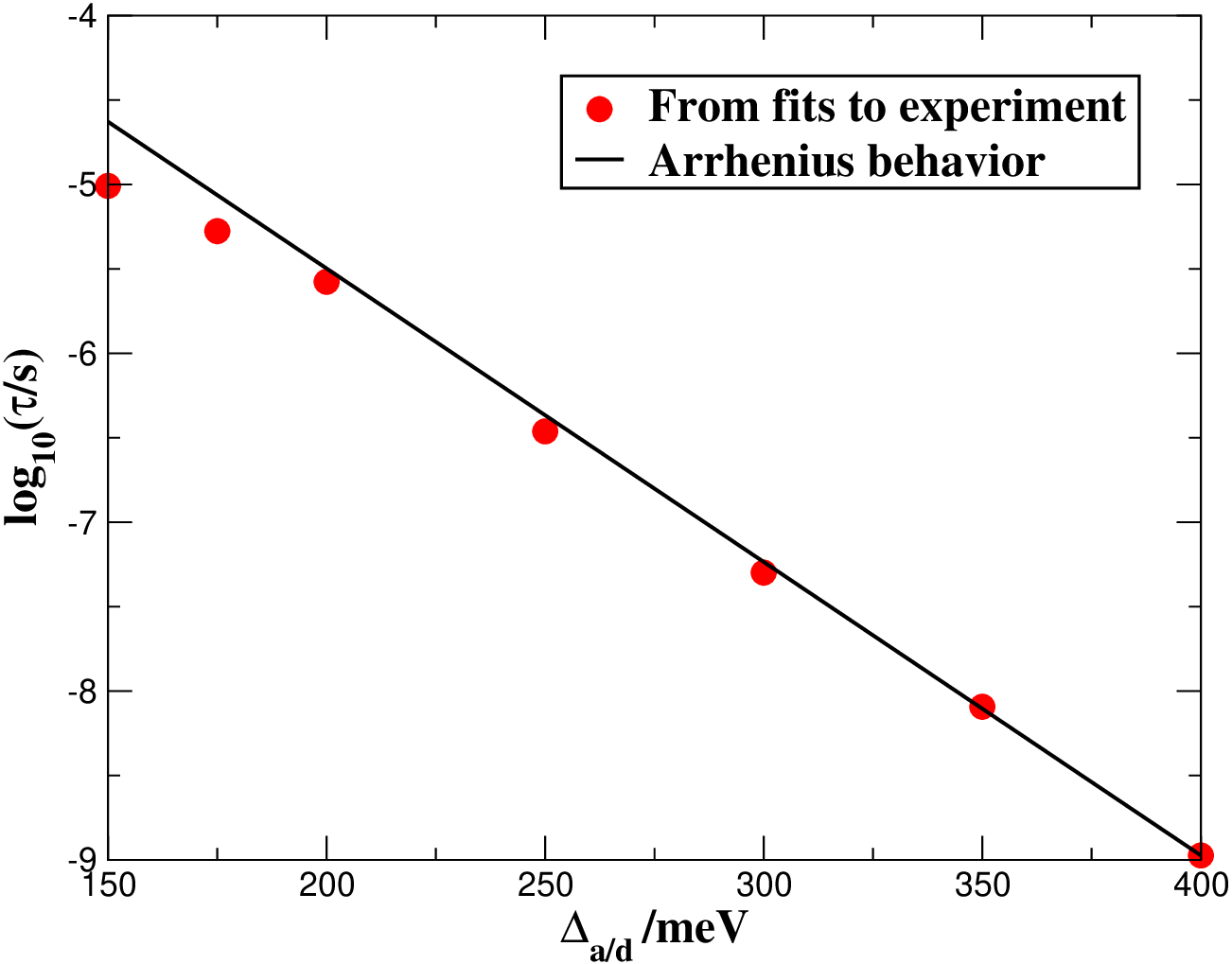}}
\caption[]{Arrhenius behavior of $\tau\approx\tau_{0}e^{-\beta\Delta_{\rm
  a/d}}$ as a function of $\Delta_{\rm a/d}$. The slope of the line
  representing the linear regression of $\log_{10}{(\tau/\rm{s})}$ for 
  $\Delta_{\rm a/d}\geq$~200~meV is $(-0.0388/\ln10)
  \; \rm{meV^{-1}}$, in very good agreement to what is expected from the
  Arrhenius equation: $\beta=0.04\; \rm{meV^{-1}}$. As $\Delta_{\rm
  a/d}$ decreases closer to the values of $\Delta_{\rm nnn}$ and
  $\Delta_{\rm nn}$, $\tau$ deviates from the Arrhenius behavior.
\label{fig:inversetime}}
\end{figure}

For $\Delta_{\rm a/d} > 200 $~meV, the physical time $\tau$
corresponding to one MCSS is simply related to $\Delta_{\rm a/d}$ 
as seen in Fig.~\ref{fig:inversetime}.
 The relationship between $\tau$ and
$\Delta_{\rm a/d}$ is related to the Arrhenius form:
\begin{equation}
\tau=\tau_{0} \exp (-\beta \Delta_{\rm a/d})\; ,
\end{equation}
or
\begin{equation}
\log_{10}{(\tau/\rm{s})} = \left( \ln (\tau_{0}/\rm{s})-\beta
\Delta_{\rm a/d}\right)/\ln{10}.
\label{eq:tau}
\end{equation}
\makebox[\textwidth][s]{%
Plotting $\log_{10}{(\tau/\rm{s})}$ vs $\Delta_{\rm a/d}$ 
results in a straight
line with a slope of} $-0.0388\; \rm{meV^{-1}}/\ln 10$ (excluding
$\Delta_{\rm a/d}=150\;{\rm and}\; 175$~meV), in very good agreement
with what is expected from Eq.~(\ref{eq:tau}) with the inverse
temperature used, $\beta=0.04\; \rm{meV^{-1}}$. \fussy We also find that,
as $\Delta_{\rm a/d}$ decreases and becomes comparable to
$\Delta_{\rm nnn}$ and $\Delta_{\rm nn}$, diffusion becomes relatively more
important in determining the overall timescale, and the dependence of
$\tau$ on $\Delta_{\rm a/d}$ deviates from Eq.~(\ref{eq:tau}). This
deviation makes differentiation among different values of $\Delta_{\rm a/d}$
easier because in this limit a change in $\Delta_{\rm a/d}$
cannot be compensated by a change in $\tau$, and the details of the 
dynamics directly influence the peak separation.
\section{Conclusions}
\label{sec:Conclusions}

By comparing with experiments simulations at potential-scan rates sufficiently
slow to produce peak separations that fall within the experimental range,
we were able to extract a physical timescale associated with the inverse 
MC attempt frequency $\tau=5.3\times10^{-6}$~s, a value much larger than
normally expected. This may be the result of
relatively flat potential minima for the adsorption process, possibly related
to reorganization of the ion hydration shells. 
Another possible reason is the assumption that 
all processes in the MC have the same attempt frequency; different processes
could, in general, have different MC attempt frequencies. Thus $\tau$ would
correspond to an effective inverse MC attempt frequency for an overall process. 

For values of the 
adsorption/desorption free-energy barrier $\Delta_{\rm a/d}$
that are large compared to the diffusion barriers ($\Delta_{\rm a/d}\geq$~200~meV),
the relationship between $\tau$ and  $\Delta_{\rm a/d}$ is consistent with
the Arrhenius law. Under such conditions the process of Br
adsorption/desorption controls the dynamics in this electrochemical
system, and the possible difference in attempt frequencies for different
processes becomes less important. For situations in which $\Delta_{\rm a/d}$
is comparable to the values of the lateral diffusion of adsorbates on the
substrate, deviations from the Arrhenius law appear. Then no process is
dominant. In conclusion we have here shown that simulations which measure 
the competition between these processes can be used to distinguish between
different $\Delta_{\rm a/d}$ values by comparison to experiments.

\section*{Acknowledgments}

We thank J.X. Wang for supplying us with the experimental data, 
S.J.~Mitchell for useful discussions, and A.P.J. Jansen for helpful
comments. This work was supported in part by NSF grant
No.~DMR-0240078 and by Florida State 
University through the School of Computational
Science and the Center for Materials Research and Technology.

\clearpage 




\end{document}